\documentclass{sigplanconf}

\usepackage{amsmath}
\usepackage{hyperref}

\begin{document}

\conferenceinfo{CAST '06}{June 5-7, 2006, Indianapolis, Indiana, USA.
} 
\copyrightyear{2006} 
\copyrightdata{Julian Harty} 

\preprintfooter{CAST Proceedings}   % 'preprint' option specified.

\title{Security testing using JUnit and Perl scripts}
\subtitle{(Reproduced in 2021 as the original publication is no longer available for you)}
\authorinfo{Julian Harty}
           {Commercetest Limited \\High Wycombe, UK}
           {julianharty@gmail.com}

\maketitle

\begin{abstract}
In this paper, I describe a recent practical experience where JUnit
was used for testing security bugs in addition to functional bugs.
Perl scripts were also used during the exploration phase. The
application being tested was mature, but insecure.
\end{abstract}

\category{D2.5}{Testing and Debugging}{}
\category{D4.9}{perl}{scripts}

\terms
Experimentation, Security, Verification.

\keywords
JUnit, Security testing.

\section{Introduction}
This presentation describes how I used a combination of short, basic scripts written in Perl, and more structured tests written in Java, using the JUnit~\cite{junit} framework to test the security of a custom file transfer program. Some details of the example have been changed to maintain confidentiality of the project, and to simplify some non-essential details, however the lessons are generally applicable to other networked software applications. 

Section 2 describes the background of the case study, Section 3 describes the exploration process and how using Perl scripts
helped to improve the effectiveness and efficiency of the testing. The rationale behind the choice of JUnit, and how the test framework was structured to perform the security testing is contained in Section 4. Section 5 contains a summary of the results of the testing, and some conclusions. Section 6 describes opportunities for further work, and includes suggestions for how readers may apply some of the lessons learnt and techniques covered in this presentation. Section 7 contains the acknowledgements, and finally Section 8 contains references and further reading.

Optionally, an additional section can be added to describe how Threat Modelling helped motivate the project management into fixing the security flaws identified during the exploration phase. This would be added between the current sections 3 \& 4. The presentation would probably take a double-session to cover the topic adequately.

\section{Background}
The case study was the result of a consultancy assignment to assess, test and trouble-shoot the performance of a 20,000 node, distributed client-server application in Europe. During the initial assessment, the file transfer programs were identified as giving ‘cause for concern’ as they underpinned a number of vital infrastructure services within the 20,000 network, including: software distribution, and collection of audit data. Failure of the file transfer sub-system would have had very serious financial, business and political implications.

While trying to analyze the behavior of the file transfer software, by entering commands manually via a software interface, miss keying of the commands caused the system to expose sensitive information. As a result, I was given permission to explore the system to find possible security flaws.

\section{Exploring using Perl scripts}

\subsection{Why scripts?}
Manual testing helped to find some security flaws with the file transfer software, however a number of reasons justified automating a number of the tests. These reasons included:
\begin{itemize}
    \item Making the tests more repeatable
    \item Simplifying the correct population and formatting of binary length fields
\end{itemize}

However we did not have the budget or time to use a commercial security-testing tool. Furthermore none of the commercial security testing tools supported the custom file transfer protocol, so the effort required to automate the tests for a commercial tool might have taken as long as doing so with homebrew scripts.

\subsection{Why Perl?}
We picked Perl for two main reasons:
\begin{itemize}
    \item We had a Perl guru on the team, who offered to help create the scripts
    \item The Perl community provided numerous add-on modules, such as mod-telnet, which reduced the effort of creating suitable tests to a minimum.
\end{itemize}

\subsection{The tests}
Approximately 6 short tests were created using Perl scripts. An initial script was simply copied and pasted before being edited to create the next test. No test program exceeded 20 lines of code, including comments, and the effort of creating a more structured, or more modular code was not justified during the exploration phase. The tests included:
\begin{itemize}
    \item Sending a correctly formatted PUT (send file)
    \item Directory attacks to determine whether sensitive files and directories could be accessed
    \item Sending messages that might cause a buffer overrun in the destination software, by ‘breaking’ the rules of the protocol.
request
\end{itemize}

\subsection{Supporting tools}
An open-source, free network analyzer called Ethereal~\cite{ethereal}, and Telnet, a commonly available system utility were used during the test automation process. Four source code analyzers were also used to help identify other potential flaws in the underlying source code of the file transfer programs.

Ethereal allowed network requests and responses to be captured and reviewed. Some of the network traffic came from a typical, working system, which were used to reverse-engineer certain aspects of the file transfer protocols. Other traffic was captured to help debug initial teething-problems when creating the first few Perl scripts, owing mainly to our lack of understanding of how to use the Net-telnet library.

Telnet provided the interface for manual testing of the file transfer programs.

\section{Using JUnit for Security Testing}
\subsection{Why automate the testing?}
As new versions of the file transfer programs were being coded and released for testing, we realized there was no structured way to test the programs. In the past the testing seemed to be limited to
manual confirmatory-testing~\footnote{Testing to see whether the software did what it should do, rather
than testing to find failures}, that required over 5000 formatted binary values in a single network request. Therefore we decided to automate the testing. As there was an incomplete, but semi-functional set of functions written in Java we decided to used this code as the basis for our main suite of test automation code.

We also wanted to prove our tests were effective i.e. that they were able to generate failures in the existing, flawed file transfer programs, before submitting the newer releases for testing. This was a variation on typical regression testing as we wanted to demonstrate that the old code failed, before testing to find out whether the new code was able to cope with such attacks. As we needed to test multiple versions of the file transfer programs,
sometimes in parallel or on multiple machines test automation seemed the most practical way to execute the tests.

\subsection{Why JUnit?}
The project team had experimented with a J2EE framework for test automation, however the framework code was complex and hard to configure. JUnit is a well-tried and respected test automation framework, particularly suited to low-level ‘developer-testing’ so we chose to use it to structure our test automation code. The developers liked being able to execute the tests from within their IDE (Eclipse) while the testers could use JUnit from the command line on test machines.

\subsection{Modifying the interface of the existing Java code}
The existing Java functions had been designed and implemented to encapsulate the inner workings of the file transfer protocol. While this might have been adequate for functional testing, the
security test cases needed to abuse the protocol e.g. in order to attempt a buffer-overflow attack. The interface had to be modified to provide the calling code (the JUnit test cases) with the ability to modify and subvert aspects of the file transfer protocol.

\subsection{Categories of test cases}
A number of types or categories of test cases were designed and
implemented. These included:
\begin{itemize}
    \item Boundary Value Analysis (BVA) of numeric fields such as buffer sizes
    \item Missing values
    \item Very large, zero and negative values for numeric fields
    \item Long strings and messages
    \item Mal-formed sequences of messages that broke the rules of the file transfer protocol
\end{itemize}

The majority of these tests were coded as separate sets of JUnit test cases e.g. one file might contain the BVA tests for the buffer length, and another the negative, zero and large integer values for a length field. However a few tests, including some for the malformed sequences of messages were left as Perl scripts as we did not have the time or need to recode these as JUnit scripts. We would have first had to create additional functions to support the PUT (send file) requests..

\section{Results and Conclusions}
\subsection{Results}
In summary, the results of the testing framework described in this paper, enabled a significant number serious security flaws to be discovered, fixed and tested in a safe, controlled environment. When new flaws were discovered, suitable tests were created in either Perl or JUnit to ensure the fixes could be tested effectively
and efficiently. Detailed results cannot be provided at this stage for various (undisclosed) reasons.

\subsection{Conclusions}
By using a combination of JUnit and Perl scripts we were able to create effective and efficient automated test cases. A number of supporting tools made the work easier.

Based on my experience of this work I suggest a similar approach may be fruitful for testing security of other types of networked applications, including web-based and XML-based software.

\section{Opportunities for further work}
This case study is based on a particular experience where many details are confidential. Therefore some details of the file transfer protocol, the security flaws found, etc. need to be obscured or obfuscated, which makes some lessons harder to transfer to interested readers. If a similar, public domain network program, particularly one that is insecure in its basic version (e.g. from an early version of the program written before security issues were prevalent) the case study could be updated to use that program as the software to be tested. Detailed results, and examples of the actual bugs found, could then be published.

\acks
My thanks to Michael Kelly who reviewed a description of this case study in another form, and to several anonymous people who helped me on the case study.

% We recommend abbrvnat bibliography style.

\bibliographystyle{abbrvnat}

% The bibliography should be embedded for final submission.

\end{document}